\documentclass[vecphys]{svmult}

\usepackage{makeidx}         
\usepackage{graphicx}        
\usepackage{multicol}        
\usepackage[bottom]{footmisc}
\usepackage{lscape}
\usepackage{mathrsfs}
\usepackage[abs]{overpic}
\usepackage{epsfig}
\usepackage{rotating}
\usepackage{fancyheadings}
\usepackage{dcolumn}
\usepackage{color}  
\usepackage{eso-pic}
\usepackage{textcomp}
\usepackage{amsmath}
\usepackage{amssymb}
\usepackage{xspace}
\usepackage{relsize}
\usepackage{pstricks}
\usepackage{pst-node}
\usepackage{floatflt}

\makeindex             

\newcommand{\ie}{i.\ e.,}					

\newcommand{\Lumi}{\ensuremath{\mathcal{L}}}			
\newcommand{\lumipb}{\mbox{pb$^{-1}$}}				
\newcommand{\lumifb}{\mbox{fb$^{-1}$}}				

\newcommand{\mum}{\mbox{$\mu$m}}				

\newcommand{\br}{\ensuremath{\mathcal{B}}}

\newcommand{\tev}{\ensuremath{\mathrm{Te\kern -0.1em V}}}
\newcommand{\gev}{\ensuremath{\mathrm{Ge\kern -0.1em V}}}	
\newcommand{\mev}{\ensuremath{\mathrm{Me\kern -0.1em V}}}	
\newcommand{\kev}{\ensuremath{\mathrm{ke\kern -0.1em V}}}	
\newcommand{\massmev}{\mbox{\mev/$c^2$}}			
\newcommand{\pgev}{\mbox{\gev/$c$}}				

\newcommand{\CP}{\ensuremath{\mathsf{CP}}}			
					
\newcommand{\pap}{\proton\antiproton}			
\newcommand{\pt}{\ensuremath{p_{\rm{T}}}}			
\newcommand{\ptb}{\ensuremath{\pt(B)}}				

\newcommand{\cdfii}{CDF II}

\newcommand{\proton}{\ensuremath{\mathrm{p}}}
\newcommand{\antiproton}{\ensuremath{\bar{\rm{p}}}}

\newcommand{\bd}{\ensuremath{B^{0}}}				
\newcommand{\bs}{\ensuremath{B^{0}_s}}				
\newcommand{\bu}{\ensuremath{B^{+}}}				

\newcommand{\bhadron}{\mbox{$b$-hadron}}			

\newcommand{\bg}{\ensuremath{B}}				
\newcommand{\bgmeson}{\mbox{$b$-meson}}				
\newcommand{\bgmesons}{\mbox{$b$-mesons}}			

\newcommand{\bn}{\ensuremath{B^{0}_{(s)}}}			
\newcommand{\bnmeson}{\mbox{$B^0_{(s)}$ meson}}			

\newcommand{\lambdab}{\ensuremath{\Lambda^0_{b}}}

\newcommand{\bmumu}{\ensuremath{\bn \rightarrow \mu^{+}\mu^{-}}}
\newcommand{\bsmumu}{\ensuremath{\bs \rightarrow \mu^{+}\mu^{-}}}
\newcommand{\bdmumu}{\ensuremath{\bd \rightarrow \mu^{+}\mu^{-}}}

\newcommand{\bhh}{\ensuremath{\bn \rightarrow h^{+}h^{'-}}}

\newcommand{\bdkpi}{\ensuremath{\bd \rightarrow K^+ \pi^-}}

\newcommand{\bskpi}{\ensuremath{\bs \rightarrow K^- \pi^+}}

\newcommand{\bspipi}{\ensuremath{\bs \rightarrow  \pi^+ \pi^-}}

\newcommand{\bdkk}{\ensuremath{\bd \rightarrow  K^+ K^-}}

\newcommand{\bsjpsiphi}{\ensuremath{\bs \rightarrow  \jpsi \phi}}

\newcommand{\lambdabppi}{\ensuremath{\lambdab \rightarrow \proton \pi^{-}}}
\newcommand{\lambdabpk}{\ensuremath{\lambdab \rightarrow \proton K^{-}}}

\newcommand{\jpsi}{\ensuremath{J/\psi}}

\newcommand{\dzero}{\ensuremath{D^{0}}}

\newcommand{\fig}[1]{Fig.\ \ref{fig:#1}}

\newcommand{\refcita}[1]{Ref.\ \cite{#1}}

\newcommand{\cita}[1]{\cite{#1}}

\newcommand{\beq}{\begin{equation}}
\newcommand{\eeq}{\end{equation}}  
\newcommand{\beqn}{\begin{eqnarray}}
\newcommand{\eeqn}{\end{eqnarray}}

\newcommand{\dedx}{\ensuremath{\it{dE/dx}}}

\newcommand{\acp}{\ensuremath{A_{\mathsf{CP}}}}
\newcommand{\acpbdkpi}{\ensuremath{\acp(\bdkpi)}}
\newcommand{\acpbskpi}{\ensuremath{\acp(\bskpi)}}

\begin{document}


\title*{Charmless $b$-hadron decays at CDF}
\author{Diego Tonelli (for the CDF Collaboration)}
\institute{Fermilab, MS 223, P.O.~Box 500 Batavia, IL 60510-500, USA, \texttt{tonel@fnal.gov}}
\maketitle

\section{Introduction}
\label{sec:1}
Measurements from the upgraded Collider Detector at the Fermilab Tevatron (\cdfii) are becoming increasingly competitive with $B$-factories 
results on \bd\ decays into charged final states, and complementary to them in corresponding \bs\ and baryon modes \cita{bhh_prl}.
In addition, the reached sensitivity to flavor-changing neutral current (FCNC) 
 \bgmeson\ decays could reveal new physics before the start-up of the Large Hadron Collider (LHC) \cita{bmumu_prl}.\par
We present recent results on these topics, from samples corresponding to time-integrated luminosities of $\int\Lumi dt\simeq 0.78$--1 \lumifb.
\textsf{C}-conjugate modes are implied throughout the text, branching fractions (\br) indicate \CP-averages, and the first (second) uncertainty 
associated to any number is statistical (systematic). Details on the \cdfii\ detector can be found elsewhere \cita{tdr}.

\section{\bhh\ decay rates}

CDF is the only experiment, to date, that has simultaneous access to \bd\ and \bs\ two-body decays into charged kaons and pions (\bhh).
 Joint study of these modes, related by flavor symmetries, may allow (partial) cancellation of hadronic unknowns 
in the extraction of quark flavor-mixing parameters.
\par We analyzed a $\int\Lumi dt\simeq 1$ \lumifb\ sample of pairs of oppositely-charged particles, used to form \bnmeson\ candidates, 
 with $p_T > 2$ \pgev\ and   $p_T(1) + p_T(2) > 5.5$ \pgev. The trigger also requires a $20^\circ < \Delta\phi < 135^\circ$ 
transverse opening-angle between tracks to reject light-quark background.  
In addition, both charged particles must originate from a transversely-displaced vertex from the beam (100 $\mu$m $< d_0 < 1$ mm), while the \bnmeson\ candidate must be produced in the primary \pap\ interaction ($d_0(B)< 140$ $\mu$m) and to travel a transverse distance $L_{xy}(B)>200~\mum$.\par
 A \bhh\ signal ($\br \approx 10^{-5}$) of about 15,000 events and signal-to-noise ratio $\rm{SNR} \simeq 0.2$ at peak is visible already after the trigger selection: 
a remarkable achievement at a hadron collider, made possible by the CDF trigger on displaced tracks \cita{svt}.\par 
In the offline analysis, an unbiased optimization further tightens the selection of track-pairs fit to a common decay-vertex. 
We use different selections, each obtained by maximizing the statistical resolution on the specific parameter to be measured (\br\ or \acp), as predicted from repeating the actual measurement on pseudo-experiments. We also exploit the discriminating power of the \bnmeson\ `isolation' and of the information provided by the 3D-view of CDF tracking, which both greatly improve signal purity. Isolation is defined as $I(B)= \ptb/[\ptb + \sum_{i} \pt(i)]$, 
where the sum runs over every other track in a cone of unit radius in $\eta-\phi$ around the \bnmeson\ flight-direction.
The $I(B) > 0.5$ requirement exploits the harder fragmentation of \bgmesons\ with respect to light-quark background. 
The 3D-view of tracking allows resolving multiple vertices along the beam direction. This halves the combinatoric background, 
with little inefficiency on signal, by removing pairs of displaced tracks from distinct, uncorrelated, heavy-flavor decays.
\par
\begin{figure}[ht]
\centering
\includegraphics[height=45mm]{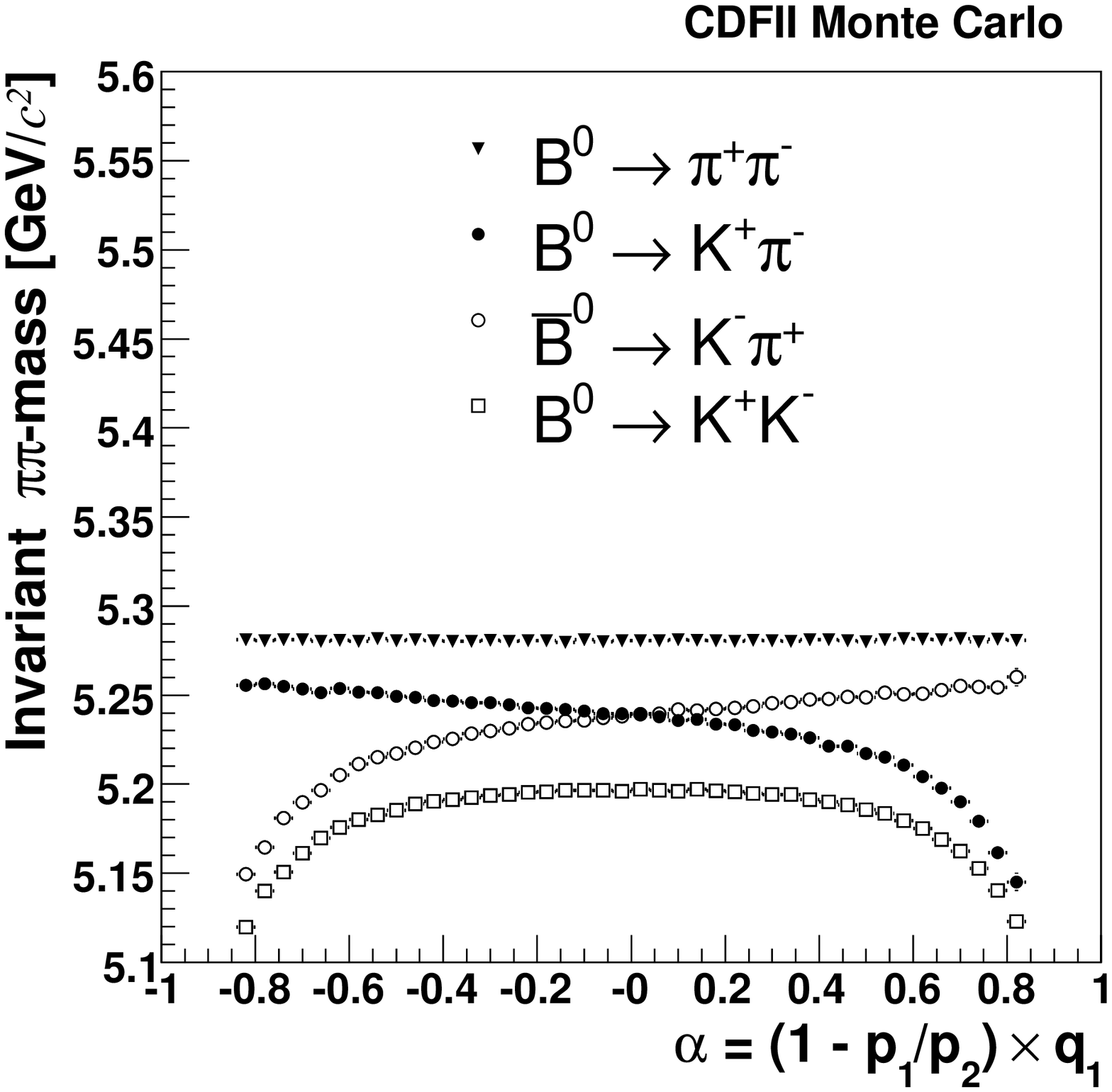}
\includegraphics[height=45mm]{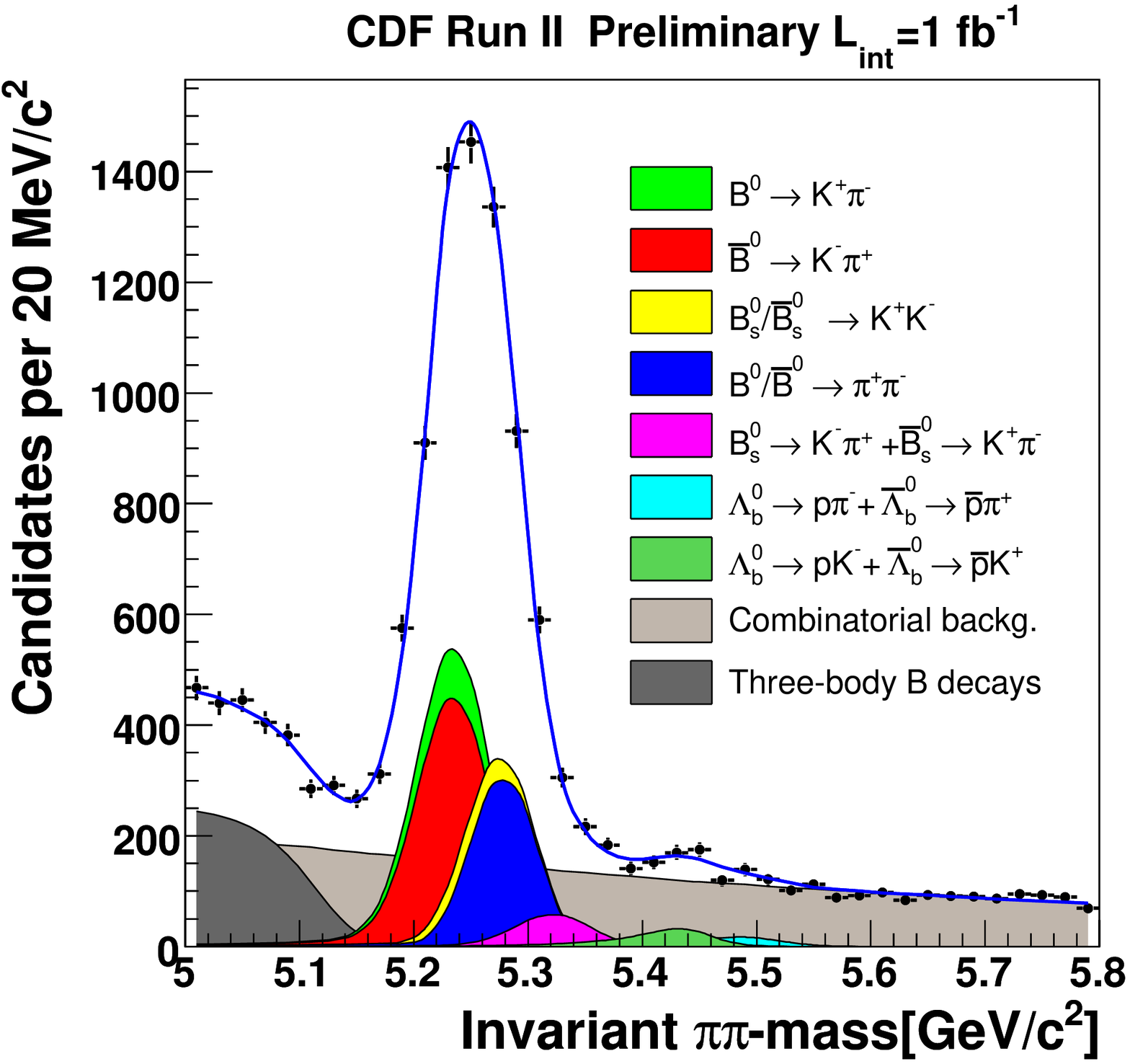}
\caption{
Invariant $\pi\pi$-mass for simulated \bd\ decays as a function of the signed ratio of momenta, $\alpha = q_{\rm{min}}(1-p_{\rm{min}}/p_{\rm{max}})$,
 where ``min'' and ``max'' refer to the magnitudes of momenta, and $q$ is the sign of the charge. Similar dependencies hold for \bs\ and \lambdab\ decays.
Invariant $\pi\pi$-mass after the offline selection with individual signal components (cumulative) and backgrounds (overlapping) 
overlaid.}  
 \label{fig:bhh}
\end{figure}

 The resulting $\pi\pi$-mass distribution  (\fig{bhh}, right) shows a clean signal, estimated by a Gaussian 
plus an exponential (combinatoric background) and an Argus-shaped (partially reconstructed \bg\ decays) fit to contain about 7,000 events,
 with standard deviation $\sigma = 39 \pm 1~\massmev$ and $\rm{SNR} \simeq 8.3$ at peak. This corresponds to a factor of 2 (40) reduction in signal (background) yield 
with respect to the trigger selection. \par
The various \bhh\ modes appear overlapping into an unresolved mass peak. Indeed, the mass and PID resolutions are
insufficient for separating them on a per-event basis. We achieved a statistical separation with a
 multivariate, unbinned likelihood-fit that uses PID information, provided by specific ionization energy loss (\dedx) in the drift chamber, 
and kinematics.\par
We exploit the kinematic differences among modes by using the correlation between masses and (signed ratios of) momenta (\fig{bhh}, left).
Mass line-shapes are accurately described accounting for the effect of final state radiation of soft photons
and non-Gaussian resolution tails. The \dedx\ is calibrated over the tracking volume and time using  
about $10^{6}$, 95\% pure, $D^{*+}\rightarrow D^0 (\to K^-\pi^+)\pi^+$ decays, where the identity of Cabibbo-favored \dzero\ decay-products is tagged 
by the strong $D^{*+}$ decay \cita{tesi_diego}. A $1.5\sigma$ separation is obtained between kaons and pions with $p>2~\pgev$.
A 10\% residual track-to-track correlation due to uncorrected common-mode \dedx\ fluctuations is included in the fit. Kinematic fit templates
 are extracted from simulation (signal) and from real mass-sidebands data (background); \dedx\ templates (signal and background) 
are extracted from the $D^0$ samples used in calibration.
\par The fitted yields reveal the first observation of \bskpi\ ($230 \pm 34 \pm 16$ events, 8$\sigma$ significance), \lambdabppi\ ($110 \pm 18\pm 16$ events, 11$\sigma$ significance), and  \lambdabpk\ ($156\pm 20\pm 11$ events, 6$\sigma$ significance) decays. After correcting for trigger, reconstruction, and selection efficiencies, we obtain the following branching fractions: $\br(\bskpi) = (5.0 \pm 0.75 \pm 1.0)\times 10^{-6}$, $\br(\bspipi)  = (5.3 \pm 3.1 \pm 4.0)\times 10^{-7}$, and $\br(\bdkk) = (3.9 \pm 1.6 \pm 1.2)\times 10^{-7}$. The extracted \CP-violating asymmetries, $\acpbdkpi = (-8.6 \pm 2.3 \pm 0.9)\%$ and $\acpbskpi = (39 \pm 15 \pm 8)\%$, can be compared for a model-independent test for presence of new physics in these decays \cita{lipkin}.
Dominant systematic uncertainties, evaluated with pseudo-experiments, include contributions from imperfect knowledge of \dedx\ shapes, isolation efficiency, combinatorial background shapes, and charge-asymmetries in background. Further details on the analysis can be found in \refcita{bhh}. 
  
\section{Search for rare FCNC $B$ meson decays}
\label{sec:bmumu}
In the standard model (SM), FCNC decays are strongly suppressed: $\mathcal{O}(10^{-9}-10^{-10})$ expected branching fractions for rare
 \bmumu\ decays are a factor $\mathcal{O}(100)$ beyond current experimental sensitivity. However, contributions from non-SM physics 
may significantly enhance these rates, making possible an observation that would be unambiguous signature of new physics.
\par We searched for \bmumu\ decays in $\int\Lumi dt\simeq 780$ \lumipb\ of data collected by the dimuon trigger. 
Offline, we require two oppositely-charged muon candidates fit to a common 
decay-vertex. We cut on the dimuon transverse momentum to reject combinatoric background, on the 3D decay-length ($\lambda$) and on its resolution 
 to reject prompt background, and on the isolation; we also require the candidate to point back to the primary vertex
 to further reduce combinatoric background and partially reconstructed \bhadron\ decays. 
This results in about 23,000 candidates, mostly due to combinatoric background.  
\par Further purity is obtained by cutting on a the likelihood-ratio (LR) based on three input observables: the isolation
 of the candidate, the decay-length probability ($e^{-ct/c\tau}$), and the `pointing' to the primary vertex
 (\ie\ the opening angle $\Delta\alpha$ between the \ptb-vector and the vector of the displacement between the \pap\ vertex and the candidate decay-vertex).  We extract the signal (background) template from simulation (mass-sidebands in data).\par
\begin{figure}[t]
\centering
\includegraphics[height=45mm]{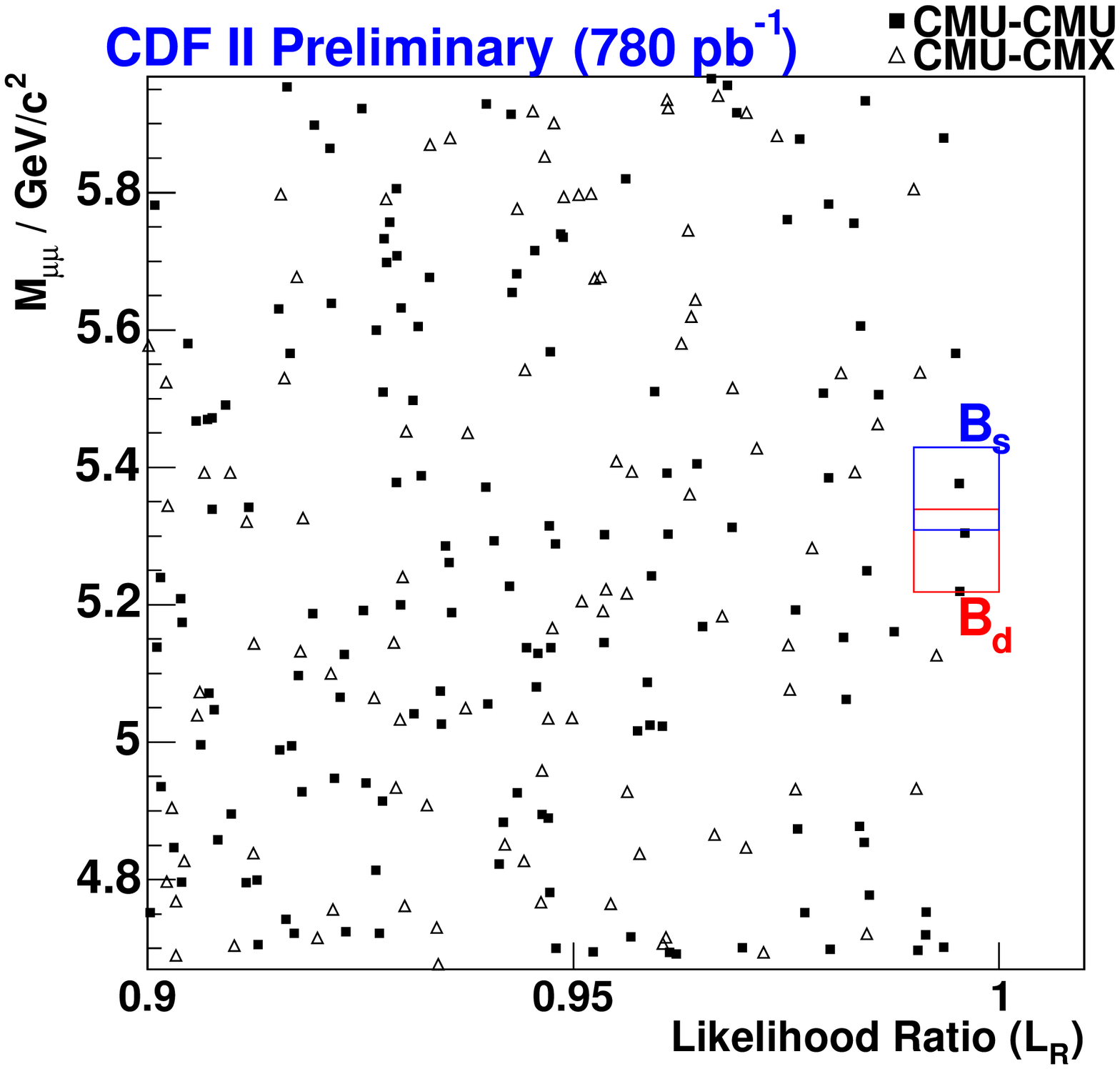}
\includegraphics[height=47mm]{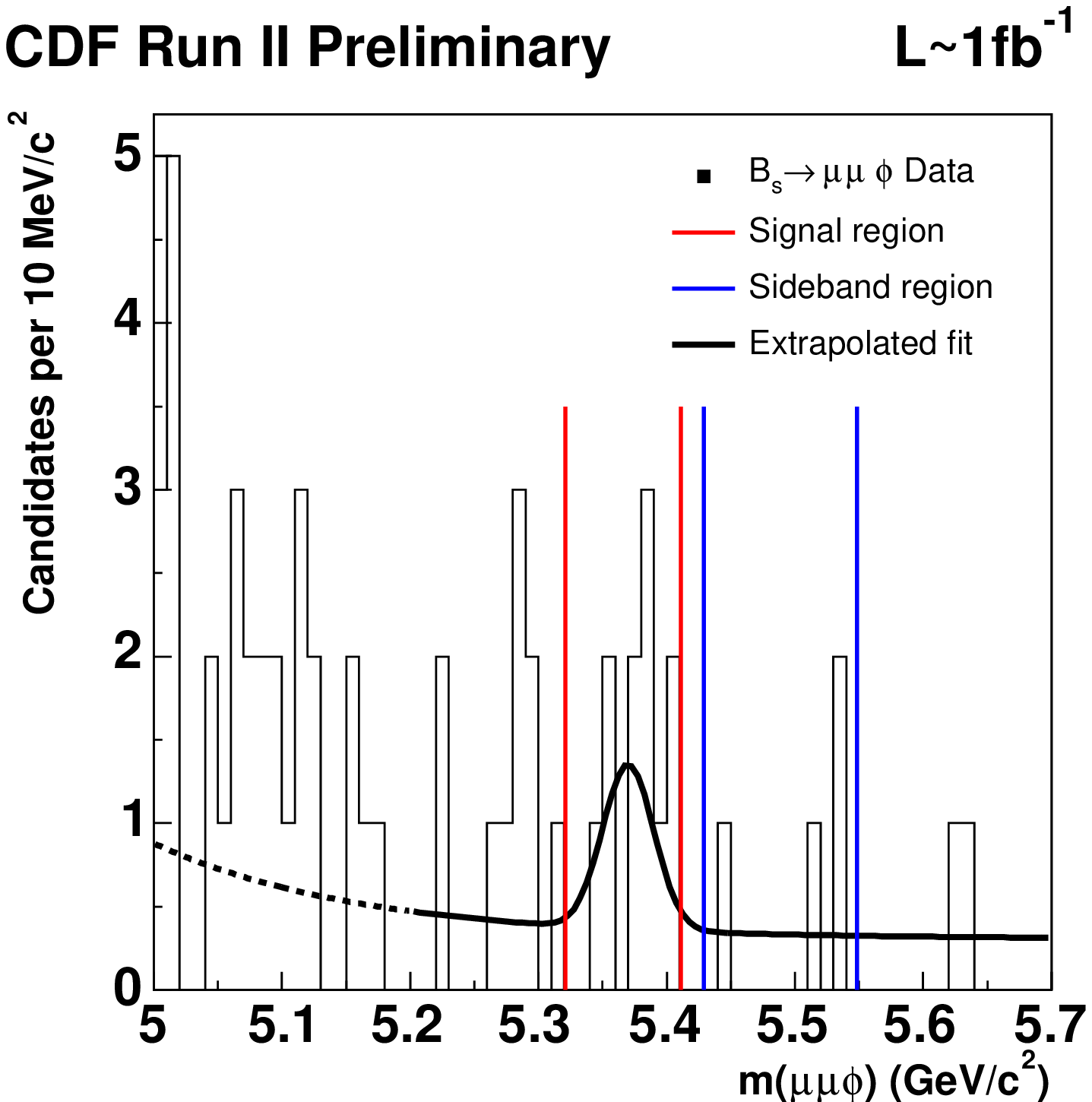}
\caption{Invariant $\mu^{+}\mu^{-}$-mass versus LR distribution (left). Both muons in the
 $|\eta|<0.6$ region (solid squares) and one in the $0.6< |\eta|<1.0$ region (open triangles). 
The \bs\ (blue box) and \bd\ (red box) signal regions are also shown.  Invariant $\mu^{+}\mu^{-}K^+K^-$-mass for events satisfying the offline selection for the $\bs\to\mu^+\mu^-\phi$ search (right).}\label{fig:bmumu_box}
\end{figure}

The \bmumu\ branching fractions are obtained by normalizing to the number of $\bu \rightarrow \jpsi (\to \mu^+\mu^-) K^+$ 
decays collected  in the same sample. The ratio of trigger acceptances between signal and normalization mode ($\simeq 25\%$) and 
the relative offline-selection efficiency ($\simeq 90\%$) are derived from simulation, the relative trigger efficiencies ($\simeq 1$) are extracted 
from unbiased data. The expected average background is obtained by extrapolating events from the mass-sidebands to the search regions.
 This estimate was checked  by comparing predicted and observed background yields in control samples such as like-sign dimuon 
candidates, and opposite-sign dimuon candidates with negative decay-length or with one muon failing the quality requirements.
 Contributions of punch-through hadrons from \bhh\ decays are also included in the estimate of total background. 
The LR cut was optimized by searching for the best \emph{a priori} expected 90\% confidence level (CL) 
upper limit on $\br(\bmumu)$. The observed event yields in two, 120 \massmev-wide search windows (to be compared with 25 \massmev\ mass-resolution) 
centered at the world average \bnmeson\ masses (\fig{bmumu_box}, left), are in agreement with the expected background events. 
A Bayesian approach that assumes a flat prior is used to estimate the following upper limits for the branching fractions: $\br(\bsmumu) < 8.0 (10) \times 10^{-8}~\rm{at}~90 (95)\%~\rm{CL}$ and $\br(\bdmumu) < 2.3 (3.0) \times 10^{-8}~\rm{at}~90 (95)\%~\rm{CL}$. 
These results improve by a factor of two previous limits and significantly reduce the allowed parameter space for a broad range of SUSY models
\cita{bmumu_theory}.
\par An analogous search is performed in 0.92~\lumifb\ for FCNC $B \to \mu^+\mu^- h$ decays, where $B= \bu,~\bd,~\mbox{or}~\bs$ and $h= K^+,~K^{*0}(\to K^+\pi^-),~\mbox{or}~\phi (\to K^+K^-)$, respectively \cita{sinead}. While $\bu$ and $\bd$ channels are already explored at the $B$-factories, the $\bs$ mode is still unobserved. 
The strategy is similar to the one used for the $\bmumu$ search: the selection is optimized by maximizing $S/\sqrt{S+B}$, where $S~(B)$ are simulated signal (real background) events. Dimuon candidates consistent with $\jpsi$ and $\psi'$ decays are removed, as those consistent with $B \to D\pi$ decays in which hadrons are misidentified as muons. The observed signal yields are obtained by counting the events in a $2\sigma$-wide window centered at the relevant
 $B$ meson mass after subtracting the background extrapolated from events in  the higher-mass sideband. The yields are normalized to the reference $B \to \jpsi h$ modes. The measured branching ratio for the $\bs$ mode, $(1.16 \pm 0.56  \pm 0.42 )\times 10^{-6}$ (different from zero at $2.4\sigma$),  
allows extraction of the most stringent limit to date: $\br(\bs \to \mu^+\mu^- \phi)/\br(\bsjpsiphi)< 2.30(2.61)\times 10^{-3}$ at 90(95)\% CL. The results for other modes, $\br(\bu \to \mu^+\mu^- K^+) = (0.60 \pm 0.15 \pm 0.04)\times 10^{-6}$ ($4.5\sigma$) and $\br(\bd \to \mu^+\mu^- K^{*0}) = (0.82 \pm 0.31 \pm 0.10)\times 10^{-6}$ ($2.9\sigma$), are consistent and competitive with $B$-factories results.

\section{Summary}

CDF continues to pursue an highly successful program in flavor physics: we obtained the first observation 
of \bskpi, \lambdabppi, \ and \lambdabpk\ decays, a competitive measurement of the \CP-violating asymmetry in \bdkpi\ decays, and the first measurement of 
the corresponding asymmetry in \bskpi\ decays. In addition, we quote the most stringent upper limits on branching fractions of rare FCNC \bmumu\ and 
$\bs \to \mu^+\mu^- \phi$ decays, that contribute to exclude a broad portion of parameter space in several SUSY models and increase the sensitivity 
to the presence of new physics before the operation of the LHC.

\printindex
\end{document}